**Abstract**
Scholars and policymakers have vigorously debated what the impact of government spending on economic growth is. Some current research and theoretical models suggest that the reaction of economic growth to the extension of government spending can be either positive or negative. This article intends to investigate the inverted-U shaped relationship between output growth and government spending (i.e., government fixed capital formation [GFCF] and government final consumption expenditure [GFCE]). Ordinary least squares (OLS) is employed as an approach to annual data for Cambodia obtained from 1971 to 2015. The result reveals that GFCF and GFCE have an inverted-U shaped relation with economic growth and that 5.40% and 7.23% are the optimal values of GFCF and GFCE, respectively. The labour growth rate and export growth rate contribute positively to the growth rate of output. This study indicates that the increasing level of government expenditure reduces the efficacy of government spending, and also helps Cambodia's policymakers to control fiscal policy more efficiently.

# Introduction

The efficacy of fiscal policy, especially government spending, has been questioned since the global crisis in 2008. In several works in the literature, public expenditure plays an essential role in facilitating economic growth (Aschauer, 1989; Farhadi, 2015; Kodongo & Ojah, 2016), economic development (Iheanacho, 2016; Molnar et al., 2006), competitiveness and other parts of economic activities (Chen & Liu, 2018; Ravn et al., 2012). Some of the empirical research works state that economic growth positively reacts to the extension of government spending (Bose et al., 2007; Das & Ghose, 2013; Gould, 1983; Kormendi & Meguire, 1986; Lee & Lin, 1994; Ram, 1986). To recover their economy from the crisis, some governments borrow money to finance their expenditure and bail out the banking industry, therefore rapidly accumulating public debt in some countries (e.g., Italy, Spain, the USA and especially Greece). A result of this is that economic growth may be harmed. For example, Greece in 2010 faced a debt crisis, which negatively affected not only its economy but also others, especially the European economy. Some scholars have found that economic growth responds negatively to an increase in government spending (Butkiewicz & Yanikkaya, 2011; Dar & AmirKhalkhali, 2002; Fölster & Henrekson, 2001; Hasnul, 2015; Landau, 1983). Additionally, this negative result can occur due to the inefficiency of public investment management, thereby leading to unproductive investment.

Government expenditure encourages economic growth as long as financing sources of the spending come from the nation's own revenues but not from a deficit (Morozumi & Veiga, 2016). A higher ratio of government expenditure to output diminishes the value of Keynesian multiplier (Barro, 1990; Chen et al., 2017). Most recent research has found an inverted-U shaped linkage between government spending and output growth (Altunc & Aydın, 2013; Chen et al., 2017; Chen & Lee, 2005; Hok et al., 2014; Makin et al., 2019; Vedder &

Gallaway, 1998). Chen and Lee (2005) used a bootstrapping model and found that the inspected types of government spending have an inverted-U shaped relation to economic growth in China. The optimal value of total government spending, public investment and government consumption equal approximately 22.8%, 7.3% and 15% of the GDP, respectively. Asimakopoulos and Karavias (2016) employed a panel generalized method of movement approach and indicated that the inverted-U shaped connection between government final consumption expenditure (GFCE) and economic growth exists in 129 countries (43 developed countries and 86 developing countries). The optimum level of GFCE as a share of GDP equals 18.04% for full sample, 17.92% for industrialized countries and 19.12% for non-industrialized countries. Hajamini and Falahi (2018) used the panel threshold approach and showed that an inverted-U shaped linkage in 14 industrialized European countries exists on GFCE and government fixed capital formation (GFCF) with optimal magnitude to be 16.63% and 2.31% of GDP, respectively. The optimal value of government spending varies according to the economy of each country, econometric methods, data set and other factors included in the regression model. Government intervention can have a positive or negative influence on economic performance. The direction of a reaction of output growth to government spending typically depends on several factors (e.g., level of government spending as a share of GDP and types of expenditure).

Cambodia was classified as a lower-middle-income developing country in 2016 (UNDP, 2018). The Cambodian government intends to maintain economic growth, thereby converging to upper-middle-income states. The global crisis in 2008 also worsened Cambodia's economy because the GDP growth rate sharply dropped from 6.7% in 2008 to 0.1% in 2009. Cambodia's public investment as a share of GDP jumped from 5.73% in 2008 to 8.20% in 2010. Government consumption as a share of GDP increased by approximately 0.71% in the same period. It is necessary to understand the impact of government spending on economic growth in Cambodia.

Figure 1 reflects plotting two types of government expenditure (GFCE and GFCF) and GDP growth rate in Cambodia. As seen in Figure 1, the connection between both types and the rate of output growth is nonlinear. The expansionary government spending (GFCF and GFCE) leads to either an acceleration in or a slowdown in Cambodia's GDP growth rate.

However, this article aims to investigate the inverted-U shaped linkage between various types of government spending (i.e., public investment [GFCF] and government consumption [GFCE]), and the growth rate of output in Cambodia and to estimate their optimal value.

The second section reviews related theories, the third section describes the specific model and method, the fourth section presents results and makes discussion and the fifth section contains the conclusions and implications for policy.

## Literature Review

Armey (1995) employs a curve (similar to the Kuznets curve and Laffer curve) to investigate an effect of public spending as a share of GDP on economic performance and names his curve the Armey curve (seen in Figure 2). The Armey curve indicates the notion of the optimal size of government spending (i.e., an inverted-U shaped connection between government expenditure and output growth). If there is no government intervention, no property rights and no rule of law to protect individuals in a society, a low level of produced output responds to a disincentive to invest and save. In the case of a small government, an increase in government spending has a tremendous positive impact on output growth. A slight increase in government spending or collective action creates a big investment incentive due to a degree of protection for private property and a reduction in trading cost in response to the improvement of infrastructure and a reliable medium of exchange. The growth-enhancing feature of public spending shrinks as the government gets larger and larger. Economic growth reaches its peak level when the marginal benefits of government spending are zero. Further government spending harms output growth because the government raises taxes to finance the expenditure or borrows money through issuing government bonds with a high interest rate. The unbalanced budget also becomes increasingly risky for productivity growth.

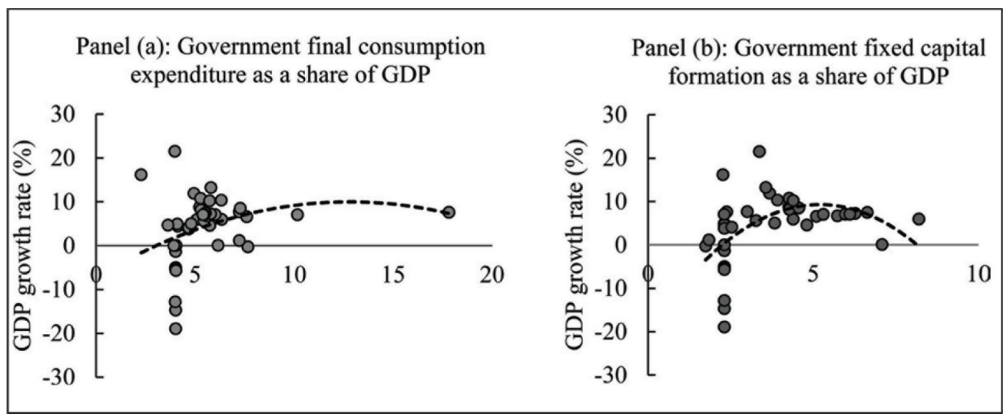

**Figure 1**. Scatter (Government Spending, GDP Growth Rate) Plot
**Source:** National Accounts Main Aggregates Database and IMF Database.
**Notes:** Each dot and dashed line represent each year and estimated line, respectively. Cambodian annual data are from 1971 to 2015. Government final consumption expenditure as a share of GDP (GFCE) and GDP growth rate are plotted in Panel (a). Panel (b) represents plotting government fixed capital formation as a share of GDP (GFCF) and GDP growth rate.

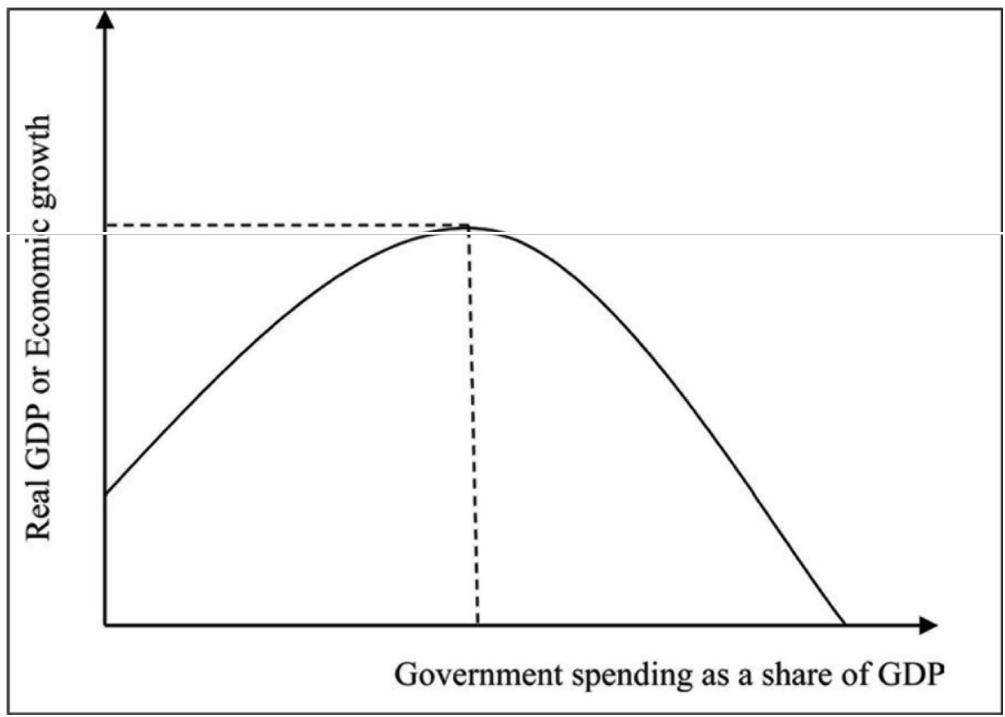

**Figure 2**. Government Expenditure as a Share of GDP and Output of Economy (Armey Curve) **Source:** Vedder and Gallaway (1998) and Chen and Lee (2005).

Barro (1990) introduces government expenditure into his endogenous growth model, blending the Ramsey model and the AK growth model and based on a fundamental assumption of constant return to scale for the production function. A flow of outputs (e.g., services of the highway, sewers and so forth) can be purchased from the private sector, and the government delivers them to households and producers free of charge and with no congestion effects. The capital of Barro's model blends physical capital with human capital improved by an increase in investment in education. In the case of a small government, an expansion of government expenditure boosts the marginal product of private sectors' capital, thereby spurring the rate of output growth. The extension of government expenditure for large government (i.e., big involvement in the expansionary fiscal policy) causes a cut in productivity growth because the government increases the tax rate to finance the spending. The optimum level of government spending as a share of GDP is when the marginal product of capital is equal to one (i.e., the natural condition for productive efficiency). If the government can maintain a balanced budget, government expenditure generates sustainable growth at the same rate. Mourmouras and Lee (1999) blend the households' utility function of Blanchard (1985) and the production function of Barro (1990) subject to finite horizons of consumers and examine the reaction of economic growth to government expenditure on infrastructure. Their result agrees with the notion of Barro (1990), but the optimal size of government expenditure on infrastructure creates a low rate of economic growth in comparison with Barro's infinite horizons.

## Methodology

*Hypothesis*

According to Barro (1990), the linkage between government expenditure as a percentage of GDP and economic growth is a quadratic function with an optimal level of government spending. This article examines two different types of government spending. The hypothesis based on Barro's observation proposes:

*H*: There is an inverted-U shaped linkage between government expenditure (i.e., GFCF and GFCE) as a share of GDP and the rate of output growth in Cambodia.

*Specific Model*

Solow (1956) and Swan (1956) highly credit production's two inputs (labour force and capital) to enhance economic growth. An accumulation of capital can be determined by government spending. Most countries are open economies nowadays; therefore, export also plays an important role in the determination of economic growth. This article tries to test the connection between government spending and output growth as follows:

$$GGDP_t = f(LAB_t, EXPO_t, GOV_t), \qquad (1)$$

Armey (1995) and Barro (1990) point out the linkage between government expenditure and output growth as a quadratic function. The regression model can be written as follows:

$$GGDP_t = \beta_0 + \beta_1 LAB_t + {} + \beta_2 EXPO_t + \beta_3 GOV_t + \beta_3 GOV_t^2 + \varepsilon_t, \qquad (2)$$

where *t* = 1971, 1972, …, 2015;
GGDP*t*: GDP growth rate of Cambodia at the time *t*;
LAB*t*: Labour force growth rate of Cambodia at the time *t*;
EXPO*t*: Growth rate of export of goods and services of Cambodia at the time *t*;
GOV*t*: Government spending as a share of GDP of Cambodia at the time *t*;

$GOV_t^2$: Square of government spending as a share of GDP of Cambodia at the time $t$; $\varepsilon t$:

Error term at the time $t$.

Total government expenditure usually is split into two major types (i.e., current expenditure and capital expenditure). Current expenditure contains GFCE and other current expenditures (transfer payment). Transfer payment can be identified as expenditure without involvement with the transition of goods and services. Capital expenditure (public investment) focuses on investment in goods and services, especially infrastructure investment (i.e., education, health, research and development, telecommunications and transport), which generates long-run benefits. Thus, GFCE and public investment as GFCF are only investigated in this study because transfer payment data are unavailable for Cambodia. Each component of total government spending is analysed separately.

## Data Collection

Cambodian data from 1971 to 2015 generate 45 observations for analysis. The list of variables is as follows:

- GFCE as a share of GDP: The general government consumes goods and services and spends money oncollective consumption services, and then this sum is divided by the GDP;
- GFCF at a constant price 2011: Disposals of produced fixed assets subtracted from the sum of acquisitions(purchase of new or second-hand assets) and specific expenditure on services adding value to nonproduced assets;
- GDP at constant price 2011: The total value of goods and services produced during a year;
- The growth rate of GDP: A percentage change of the total value of goods and services produced in anation;
- The growth rate of export of goods and services: A percentage change of the value of goods and servicessold to the rest of the world;
- The population growth rate: A percentage change of people currently living in a country.

Three primary sources report the data of variables mentioned earlier.

- The Investment and Capital Stock Dataset of IMF offers data for GDP and GFCF at a constant price 2011through this link: https://www.imf.org/external/np/fad/publicinvestment/
- World Bank Database provides data of population growth rate at this link: https://data.worldbank.org/country/cambodia?view=chart
- United Nations Statistics Division's National Accounts Main Aggregates Database. The link to get the dataof the rest of the aforementioned variables is as follows: https://unstats.un.org/unsd/snaama/dnlList.asp

The transformation made to obtain the independent variables for the regression can be explained as follows:

- Population growth rate can be used to measure the labour force growth rate;
- GFCF at a constant price 2011 divided by GDP at a constant price 2011 equals GFCF as a share ofGDP.

STATA 15.1 was the software used to process the data analysis in this study.

## Ordinary Least Squares

Engle-Granger approach (Engle & Granger, 1987) or Johansen's multivariate maximum likelihood approach (Johansen, 1988; Johansen & Juselius, 1990) for co-integration demands all of the variables (i.e., explained and explanatory variables) to be integrated to order one I(1). Autoregressive distributed lags (ARDL) bound approach (Pesaran & Shin, 1998; Pesaran et al., 2001) requires explained variable as order one of integration I(1), but predictors can be pure order zero I(0), absolute order one I(1) or mixed orders (i.e., I(0) and I(1)) of integration. Therefore, these co-integration approaches can be applied if the dependent variable is integrated to order one I(1). In the case of all variables (dependent and independent variables) to be stationary at the level I(0), ordinary least squares (OLS) as the classical method of regression modelling can be applied for the time-series data analysis. The OLS estimate based on minimizing sum square of residuals is so-called BLUE (Best Linear Unbiased Estimate). The good-fit model is subjected to the value of $R$-squared ($R^2$). If the value of $R^2$ is high, it can be regarded as a good model. The error term (residuals) estimated by OLS has to be assumed to be a white-noise (homoscedasticity—constant variance, normal distribution—zero mean and no autocorrelation).

*Calculation of the Optimum Value of Government Spending*

The optimum level of government expenditure is calculated by taking the partial derivative of GGDP (Equation [2]) with respect to *GOV* and setting it equal to zero:

$$\frac{\partial GGDP}{\partial GOV} = \beta_3 + 2\beta_4 GOV = 0 \Rightarrow GOV = -\frac{\beta_3}{2\beta_4}, \beta_3 > 0, \beta_4 < 0, \quad (3)$$

## Results and Discussion

*Estimation*

According to econometric literature of time series, the estimation with non-stationary variables produces a spurious result of the regression (Granger & Newbold, 1973); due to this, it is necessary to conduct the unitroot test, which is used to check that time-series data include a deterministic or a stochastic trend while those series transform from non-stationarity into stationarity (Kirchgässner et al., 2013). The Augmented Dickey– Fuller (ADF) test (Dickey & Fuller, 1979) as a well-known test of a unit root in time series is used to check differencing order, which leads to stationary data. The Bayesian Information Criterion (BIC) developed by Schwarz (1978) is employed to select an optimal number of lags. The null hypothesis of this test proposes a unit root or non-stationarity. The result of ADF test presented in Table 1 indicates that the dependent variable (GGDP) and predictors (LAB, EXPO, *GFCF*, *GFCF*$^2$, *GFCE* and *GFCE*$^2$) are stationary at order zero I(0). Thus, OLS is applied to estimate the connection between explained and explanatory variables.

| Test | Augmented Dicky–Fuller (ADF) | |
| --- | --- | --- |
| | $X_i$ | $\Delta X_i$ |
| GGDP | −2.521*** | |
| LAB | −2.880*** | |
| EXPO | −4.856*** | |
| GFCF | −1.331* | |
| GFCF² | −1.642* | |
| GFCE | −3.155*** | |
| GFCE² | −3.683*** | |

Table 2 displays the result of OLS analysis. The first model (Model I) for GFCF and the second model (Model II) for GFCF provide an $R^2$ value of 60.44% and 49.85%, respectively. The coefficients of explanatory variables in both models are statistically significant at 5% level. The healthy economic growth responds to the improvement of the growth rate of the labour force ($\beta_1 \rangle 0$) because more labour generates more production in the economy. An increase in the export growth rate significantly and positively influences the growth rate of output ($\beta_2 \rangle 0$). It reflects the more significant gains from international trade, thereby promoting saving, investment and economic performance in the country. The GFCF and GFCE's hypothesis, an inverted-U shaped relation with economic growth, is not rejected. The optimal value of GFCF and GFCE is estimated to be approximately 5.40% and 7.23%, respectively. The influence of government expenditure on economic growth shrinks while steadily increasing the value of government expenditure as a share of GDP. The government expenditure financed by raising taxes and taking out loans might drive down private investment due to creating more disincentives. The growth in public investment (GFCF) above the optimal level becomes unproductive because the allocation of this government investment might finance some inefficient projects. If GFCE passes the optimal level, there might be bureaucracy and centralization, which stifle creativity in the private and public sectors. The entire economy can be harmed by reducing the scope of creativity and creating more inefficiency.

| GGDP | Model I (GFCF) | | Model II (GFCE) | |
| --- | --- | --- | --- | --- |
| | Coefficient | SE | Coefficient | SE |
| $\beta_1$ | 1.233** | 0.478 | 1.134** | 0.554 |
| $\beta_2$ | 0.103*** | 0.027 | 0.175*** | 0.040 |
| $\beta_3$ | 9.155*** | 2.517 | 2.820** | 1.358 |
| $\beta_4$ | −0.848*** | 0.281 | −0.195** | 0.072 |
| $\beta_0$ | −19.819*** | 4.850 | −9.583* | 5.074 |
| $R^2$ | 0.6044 | | 0.4985 | |
| Adjusted $R^2$ | 0.5648 | | 0.4484 | |
| Root MSE | 5.0141 | | 5.6451 | |

## Diagnostic Tests

Diagnostic tests are required to check whether the residual (error term) of OLS meets the own essential three assumptions. The Breusch–Godfrey test introduced by Breusch (1978) and Godfrey (1978) relies on Lagrange multiplier (LM) test statistic and checks the autocorrelation (serial correlation) of the residuals. The Breusch–Godfrey test's null hypothesis proposes no autocorrelation. White (1980) introduced a heteroscedasticity-consistent variance estimator of the variance matrix, called White's test, to check the heteroscedasticity of the variance of residual. The null hypothesis of this White's test suggests no heteroscedasticity. The Jarque–Bera test developed by Jarque and Bera (1987) joins between skewness and kurtosis. This test relies on asymptotic standard error without correlation for sample size. The null hypothesis of the Jarque–Bera test suggests a normal distribution. Table 3 shows that the null hypothesis of Breush–Godfrey LM test, White's test and

Jarque–Bera test is not rejected at 1% significance level. There is normality, no serial correlation and no heteroscedasticity for the residual of OLS.

| $\varepsilon_t$ | Model I | Model II |
|---|---|---|
| | Chi² | Chi² |
| Breusch–Godfrey LM test | 4.805 | 8.198 |
| White's test | 8.84 | 15.46 |
| Jarque–Bera test | 6.93 | 8.45 |

*Stability Test*

The robustness of the models describes the regression model's parameter stability confirmed by the cumulative sum test. The cumulative sum test subjected to recursive residuals and proposed in Brown et al. (1975) is designed to detect the parameters' instability (Ploberger & Krämer, 1992). No structural breaks (constant regression coefficients over time) are proposed as the null hypothesis of the cumulative sum test. The results are presented in Table 4. For Model I and Model II, the null hypothesis of the test is accepted at 1% level of significance. The convergence of estimated long-run parameters to the zero means exists on both models. Model I and Model II, therefore, are stable and consistent models.

| Model | Model I | Model II |
|---|---|---|
| Test statistic | 0.343 | 0.737 |
| Critical value 1% | 1.143 | 1.143 |
| Critical value 5% | 0.947 | 0.947 |
| Critical value 10% | 0.850 | 0.850 |

If the test statistic is smaller than critical value, the null hypothesis of the test is not rejected.

Figure 3 shows the CUSUM of Model I and Model II. The solid line in Panel (a) and (b) is located in the shaded area, so these two models have stability and consistency.

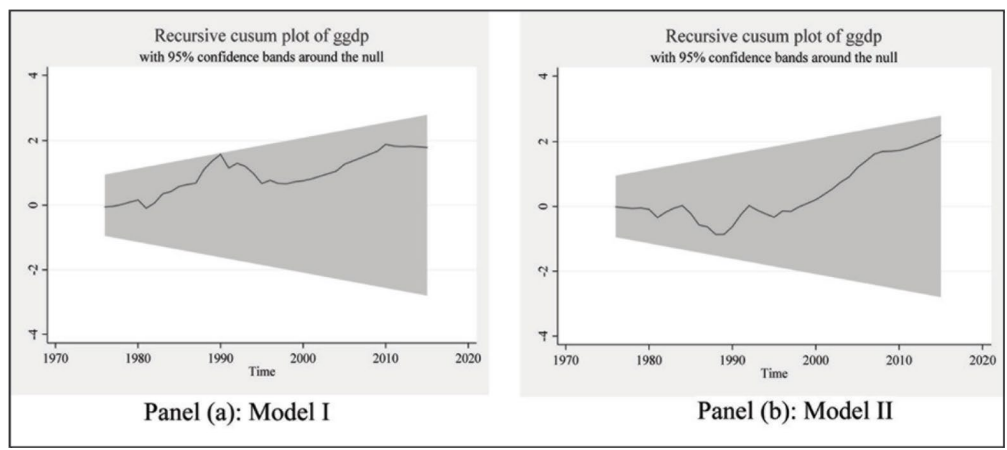

Panel (a): Model I    Panel (b): Model II

Figure 3. CUSUM for Two Models **Source:** Authors' estimation.

*Robustness Test*

The robustness of regression results of government spending (i.e., government investment and consumption) is presented in this section. The regression model takes into account more specifications (e.g., dummy variables) to shock it and is also analysed with second-degree polynomial regression.

The ADF test as a basic test of unit root is criticized for not incorporate structural breaks in time-series data, thereby producing a misleading conclusion (Glynn et al., 2007). Cambodia's history is burdened by war, genocide and occupation, times during which economic conditions are different than in peacetime. Thus, our dependent variable can be tested to find whether structural breaks appear in time-series data of the regressand. The Zivot–Andrews test developed by Zivot and Andrews (1992) incorporates unknown structural breaks in intercept, trend and both. The null hypothesis of this test suggests that time-series data are non-stationary (unit root). An alternative hypothesis is trend-stationary with a single break. The results presented in Table 5 indicate that the null hypothesis is rejected at all levels of significance, so structural breaks should be included in the regression model.

| Test | Break of Intercept | | Break of Trend | | Break of Intercept and Trend | |
|---|---|---|---|---|---|---|
| | $X_i$ | $\Delta X_i$ | $X_i$ | $\Delta X_i$ | $X_i$ | $\Delta X_i$ |
| GGDP | −6.008*** | | −6.175*** | | −6.546*** | |

Cambodia's history showed that there have been a few shocks, which affect economic conditions. Four dummy variables, therefore, are incorporated in regression. In 1973, Cambodia started a civil war between the Khmer Rouge's army led by Pol Pot and the Khmer Republic government's army with the USA's assistance led by Lon Nol. This war negatively influenced Cambodia's economy. The first dummy variable (du1) is introduced in our model. The year 1973 is given value 1, and the rest of the years are zero.

Cambodia also faced political unsettlement in 1989, thereby suddenly worsening Cambodia's economy. Our regression analysis also takes into account the second dummy variable (du2) of this political instability. The year 1989 is given value 1, and all other years are zero.

During 1994–1995, Cambodia faced political uncertainty because the Cambodia People's Party (CPP) leaders refused to accept the election outcome. The disagreement about the 1993 national election result spun out political turmoil and led to a political impasse during 1994–1995. This period is introduced as a structural break as the third dummy variable (du3). The year 1994 or 1995 is given value 1, and the rest of the years are zero.

The Asian financial crisis in 1997 started in Thailand and also contributed negatively to Cambodia's economy because they are neighbouring countries and trading partners. The fourth dummy variable (du4) denotes a structural break due to the Asian financial crisis in 1997. The value one represents the year 1997, and other years are zero.

These dummy variables are defined as follows:

$$du1 = \begin{cases} 1 \ if \ t = 1973 \\ 0 \ if \ t = other \ years \end{cases}, \quad du2 \begin{cases} 1 \ if \ t = 1989 \\ 0 \ if \ t = other \ years \end{cases},$$

$$du3 = \begin{cases} 1 \ if \ t = 1994, 1995 \\ 0 \ if \ t = other \ years \end{cases}, \quad du4 = \begin{cases} 1 \ if \ t = 1997 \\ 0 \ if \ t = other \ years \end{cases}. \tag{4}$$

There is a substantial correlation between a government's spending (i.e., government investment and consumption) and its power. Theoretical literature about the linkage between government expenditure and economic growth suggests that their relationship is a quadratic function. The second-degree polynomials of independent variables (i.e., public investment and government purchasing) are proposed in this analysis. The orthogonal polynomial terms generated by the Christoffel–Darboux recurrence formula (Abramovitz & Stegun, 1972) meets the property (i.e., quadratic trend without the constant). Equation (2) can be rewritten with the orthogonal polynomial terms of regressors (i.e., government investment and consumption) as follows:

$$GGDP_t = \beta_0 + \beta_1 LAB_t + \beta_2 EXPO_t + a_1 PGFCF1_t + a_2 PGFCF2_t + \alpha_3 PGFCE1 \\ + \alpha_4 PGFCE2_t + \alpha_5 du1 + \alpha_6 du2 + \alpha_7 du3 + \alpha_8 du4 + \varepsilon_t, \tag{5}$$

where $t$ = 1971, 1972, …, 2015;
$GGDP_t$: GDP growth rate of Cambodia at time $t$;
$LAB_t$: labour force growth rate of Cambodia at time $t$;
$EXPO_t$: growth rate of export of goods and services of Cambodia at time $t$;
$PGFCF1_t$: first degree of an orthogonal polynomial of government investment as a share of GDP of Cambodia at time $t$;
$PGFCF2_t$: second degree of an orthogonal polynomial of government investment as a share of GDP of Cambodia at time $t$;
$PGFCF1_t$: first degree of an orthogonal polynomial of government consumption as a share of GDP of Cambodia at time $t$;
$PGFCF2_t$: second degree of an orthogonal polynomial of government consumption as a share of GDP of Cambodia at time $t$; du1: dummy variable of Cambodia's civil war in 1973;
du2: dummy variable of Cambodia's political instability in 1989; du3: dummy variable of Cambodia's political instability during 1994–1995; du4: dummy variable of the Asian financial crisis 1997; $\varepsilon_t$: error term at time $t$.

The results presented in Table 6 show that the second-degree orthogonal polynomial regression provides $R^2$ (83.30%) and root mean square error (3.5337). All of the explanatory variables are statistically significant. The improvement of the labour force or growth rate of exports stimulates Cambodia's economic growth. Some shocks (i.e., a civil war in 1973, political deadlock in 1989, political instability during 1994–1995 and the Asian financial crisis in 1997) in Cambodia's history slowed down its economic growth because these shocks negatively affect household behaviour regarding expenditure and investment in Cambodia. The quadratic response to government investment has an optimal value at $PGFCF$ (orthogonal polynomial of $GFCF$) = –$a_1/(2a_2)$ = 0.43, which was approximately 5.20% on the original government investment (GFCF) scale. The inverted-U shaped relationship between government consumption and economic growth exists. The optimal level of $PGFCE$ (orthogonal polynomial of GFCE) was –$a_3/(2a_4)$ = –0.55, which was approximately 6.45% on the original government consumption (GFCE) scale. This optimal level of GFCF and GFCE is slightly lower than the optimal value from Model I and Model II.

| GGDP | Coefficient | Standard Error |
|---|---|---|
| LAB | 1.390*** | 0.353 |
| EXPO | 0.195*** | 0.031 |
| PGFCF1 | 1.427** | 0.682 |
| PGFCF2 | −1.657*** | 0.581 |
| PGFCE1 | −2.486*** | 0.703 |
| PGFCE2 | −2.264** | 0.837 |
| du1 | −15.653*** | 3.745 |
| du2 | −11.456** | 4.565 |
| du3 | −10.871*** | 3.010 |
| du4 | −6.662* | 3.714 |
| Constant | −0.588 | 0.819 |
| $R^2$ | 0.8330 | |
| Adjusted $R^2$ | 0.7838 | |
| Root MSE | 3.5337 | |

*Discussion*

The finding of this study agrees with the explanations of Barro (1990), Armey (1995) and Mourmouras and Lee (1999) about the existence of an inverted-U shaped connection between government spending and economic growth. The level of government expenditure determines whether there is a positive or negative impact, as illustrated by Keynesian theory and Neoclassical theory, respectively. A rise in government spending below the optimal level improves the investment environment, employment, consumption and therefore the economy as a whole. If it is over the threshold level, there is harm to economic performance because government spending financed by raising taxes and borrowing leads to less incentive to household consumption and investment. This finding is in line with the studies of Vedder and Gallaway (1998), Chobanov and Mladenova (2009) and Hok et al. (2014); however, they use total government expenditure as a share of GDP and various estimation methods. This result is also consistent with the outcomes of Chen and Lee (2005), Asimakopoulos and Karavias (2016) and Hajamini and Falahi (2018), who also investigated the influence of government spending's two types (e.g., GFCF and GFCE) on output growth, although these studies provide the various threshold level (optimal value).

The optimal level of GFCE calculated in this study is approximately 7.23%, which is lower than the 18.04%, 16% and 15% yielded in the studies conducted by Asimakopoulos and Karavias (2016), Chiou-Wei et al. (2010) in the case of Taiwan, and Chen and Lee (2005), respectively. The threshold level of GFCE calculated by Chiou-Wei et al. (2010) in the case of South Korea and Thailand is also higher (11%) than the optimal value in this study.

The optimal value of GFCF calculated in this study equals approximately 5.40%. This optimum value is higher than the threshold value (2.31%) reported by Hajamini and Falahi (2018), but lower than 7.3% estimated by Chen and Lee (2005) and the 13% by Davies (2009). The optimal level of GFCE and GFCF is different from other findings owing to Cambodia's historical data, economic situation, distinctive methods, and economic and social factors included in the model. The optimal value of government spending may be heterogeneous across countries. A large government finances its expenditures through taxation and allocates more spending into unproductive projects than a small government, thereby leading to the optimal level in developed countries being lower than in developing countries (Asimakopoulos & Karavias, 2016; Gray et al., 2007).

## Conclusions and Policy Implication

### *Conclusions*

Public policy has fascinated policymakers since the global crisis of 2008. Scholars have found inconsistent results, with either positive or negative impacts of government spending on output growth. The current literature also points out the non-linear relationship between them. This article explores whether Barro (1990)'s idea about government spending is valid for components (e.g., GFCF and GFCE) of total government expenditure. The Cambodia annual data from 1971 to 2015 are collected. OLS is used to estimate coefficients of each explanatory variable.

The result shows that a rise in the export growth rate or labour growth significantly spurs the growth rate of output in Cambodia. Remarkably, GFCF and GFCE meet Barro (1990)'s idea about an inverted-U relation with economic growth. Over the threshold level, the extension of GFCF and GFCE has detrimental effects on output growth due to a rise in taxes and a crowding-out effect (i.e., a reduction in investment and consumption or the elimination of private sector's spending reacts to the improvement of public spending). The optimal value of GFCF and GFCE as a share of GDP was estimated to be 5.40% and 7.23%, respectively. This outcome gives both scholars and policymakers a benefit (i.e., managing the government expenditure efficiently to get an improvement of economic growth in Cambodia). This study also contributes to a theoretical part through the existence of Keynesian or Neoclassical concept governed by the level of government expenditure.

### *Limitation*

This study has limited data. The Cambodia historical data of GFCF and GFCE from 1971 to 1986 seem unchanged. Probably some of the data are not real but were obtained through the United Nations and IMF's estimation because Cambodia was involved in a civil war during this period. The quadratic function of government spending is used in this study. The calculated optimal level of GFCF and GFCE in this study, therefore, might be above or below the real optimum level. Although Próchniak (2011) and many others suggest that a combination of demand-side and supply-side factors dictate economic growth. However, the model in this study cannot capture all determinants from both sides, especially some of the essential supplyside factors (e.g., human capital and institutional environment). The main reason is that data of these factors are unavailable or limited for Cambodia.

*Policy Implications*

The Cambodian government still holds a Keynesian perspective (i.e., the extension of government spending improves economic performance). The government has reformed the public sector (e.g., increasing the wage bill for public employees since 2011) and the education system (e.g., secondary school exam since 2014 and improvement of the curriculum). Notably, public expenditure on education as a share of GDP from 2007 to 2016 increased by approximately two-thirds (World Bank, 2017). The investment projects carried out by the government are a road network improvement project (70 million USD), a provincial water supply and sanitation project (119.3 million USD) and the Tonle Sap Poverty Reduction and Smallholder Development Project (66 million USD) (ADB, 2018).

Nevertheless, this study also pointed out the optimal level of government spending in Cambodia. According to the findings of this study, an increase in government spending above the optimal value leads to a slowdown in economic growth. The government should adjust the level of its expenditures and save some amount of money to balance the budget or to finance productive categories of spending.

The actual GFCF as a share of GDP in 2015 equals 5.30%, smaller than the 5.40% estimated in this article as the optimum value of GFCF. A slight increase in GFCF, productive investment, drives economic growth in Cambodia. The GFCE of 5.40% in 2015 has not yet exceeded the calculated optimal level (7.23%). Thus, the Cambodian government can apply an expansionary public policy to encourage the economy.

*Further Studies*

To compensate for limited data, quarterly (Makin & Ratnasiri, 2015) or semi-annual data could be used to expand the number of observations. Alternatively, increasing the number of data can be manipulated by using a panel approach, especially with CLMV (Cambodia, Laos, Myanmar and Vietnam) countries, because these countries grouped into the sub-region within the ASEAN region have a similar economy. Undertaking further studies is necessary to confirm that Barro (1990)'s perspective is valid for government expenditure on agriculture, education, health, military, research and development (R&D) and transport. If an inverted-U relation appears in these parts of government spending and if an optimal level can be determined, the level of these expenditures might be controlled more efficiently to contribute positively to economic growth.